\documentclass[superscriptaddress,twocolumn,prl,showpacs]{revtex4}
\usepackage[dvips]{graphicx}
\usepackage{amsmath}
\usepackage{bm}
\usepackage{braket}

\newcommand{\beq}{\begin{equation}}
\newcommand{\eeq}{\end{equation}}
\newcommand{\bea}{\begin{eqnarray}}
\newcommand{\eea}{\end{eqnarray}}
\newcommand{\nn}{\nonumber\\}
\newcommand{\rvec}{{\bf r}}
\newcommand{\Rvec}{{\bf R}}
\newcommand{\Avec}{{\bf A}}
\newcommand{\Bvec}{{\bf B}}
\newcommand{\pvec}{{\bf p}}

\newcommand{\vvec}{{\bf v}}
\newcommand{\kvec}{{\bf k}}
\newcommand{\Lvec}{{\bf L}}
\newcommand{\Mvec}{{\bf M}}
\newcommand{\Svec}{{\bf S}}
\newcommand{\evec}{{\bf e}}
\newcommand{\sigmavec}{\ensuremath{\bm{\sigma}}}

\newcommand{\Hcal}{\ensuremath{\mathcal{H}}}

\newcommand{\VRNL}{V_{\bf R}^\mathrm{NL}}
\newcommand{\FRNL}{F_{\bf R}^\mathrm{NL}}
\newcommand{\ERNL}{{\bf E}_{\bf R}^\mathrm{NL}}

\newcommand{\ptilde}{\ensuremath{\widetilde{p}}}
\newcommand{\eq}[1]{Eq.~(\ref{#1})}

\begin{document}
\title{First-principles theory of the orbital magnetization}
\author{Davide Ceresoli}
\affiliation{Department of Materials Science and Engineering,
Massachusetts Institute of Technology (MIT), 77 Massachusetts Avenue,
Cambridge, Massachussets 02139-4307, USA}
\author{Uwe Gerstmann}
\affiliation{IMPMC, CNRS, IPGP, Universit\'e Paris 6, Paris 7, 140 rue
de Lourmel, F-75015 Paris, France}
\author{Ari P. Seitsonen}
\affiliation{IMPMC, CNRS, IPGP, Universit\'e Paris 6, Paris 7, 140 rue
de Lourmel, F-75015 Paris, France}
\author{Francesco Mauri}
\affiliation{IMPMC, CNRS, IPGP, Universit\'e Paris 6, Paris 7, 140 rue
de Lourmel, F-75015 Paris, France}
\date{\today}

\begin{abstract}
Within density functional theory we compute the orbital magnetization
for periodic systems evaluating a recently discovered Berry-phase
formula. For the ferromagnetic metals Fe, Co, and Ni we explicitly
calculate the contribution of the interstitial regions neglected so far
in literature. We also use the orbital magnetization to compute the EPR
$g$-tensor in paramagnetic systems.  Here the new method can also be
applied in cases where linear response (LR) theory fails, e.g. radicals
and defects with an orbital-degenerate ground-state or those containing
heavy atoms.
\end{abstract}
\pacs{71.15.-m, 71.15.Mb, 75.20.-g, 76.30.-v}
\maketitle

The electric polarization and the orbital magnetization are well known
textbook topics in electromagnetism and solid state physics. While
it is easy to compute their derivatives in an extended system, the
electric polarization and the orbital magnetization themselves are not
easy to formulate in the thermodynamic limit, due to the unboundedness
of the position operator. The problem of the electric polarization
has been solved in the '90s by the Modern Theory of Polarization
(MTP)~\cite{MTP,resta-RMP}, which relates the electric polarization to
the Berry phase of the electrons.  A corresponding formula for the orbital
magnetization has been found very recently~\cite{orbmagn,orbmagn-others}
showing that this genuine bulk quantity can be evaluated from the ground
state Bloch wavefunctions of the periodic system.
Since the discovery of the MTP, a wealth of papers have appeared reporting
its successful applications to first principles calculations of dielectric
and piezoelectric properties~\cite{resta-RMP}.  On the other hand, {\em
ab-initio} calculations of the orbital magnetization via the Berry phase
formula have not been reported in literature yet, except than for simple
tight-binding lattice models.

The origin of the orbital magnetization in molecules and solids is
time-reversal breaking caused by e.g. spin-orbit (SO) coupling.
In ferromagnetic materials the orbital magnetization is a not
negligible contribution to the total magnetization.  Several
papers in literature~\cite{wu01,sharma07} showed that the orbital
magnetic moment of simple ferromagnetic metals (Fe, Co and Ni)
is strongly underestimated within density functional theory (DFT)
if using the local density approximation (LDA) or generalized gradient
approximations (GGA).  Empirical corrections like the orbital polarization
(OP)~\cite{trygg95} have been thus employed to obtain a better agreement
with the experimental values.  Nevertheless it remains an interesting
question if e.g.~functionals beyond LDA/GGA would be able to describe
the orbital magnetization correctly~\cite{sharma07}.  All previous {\em
ab-initio} calculations have been however carried out in the muffin tin
(MT) approximation, i.e. computing the orbital magnetization only in a
spherical region centered on the atoms, neglecting the contribution of
the interstitial region.

In this letter, we present first principles DFT calculations of the
orbital magnetization by evaluating the recently discovered Berry phase
formula~\cite{orbmagn,orbmagn-others}.
For the ferromagnetic phases of Fe, Co, and Ni we show that the
interstitial regions contribute by up to 50\% to the orbital magnetic
moments.  So far neglected in the literature these contributions are
thus shown to be one source for underestimated {\em ab-initio} values.
Furthermore we make use of a relationship between the orbital
magnetization and the electronic $g$-tensor that can be measured in
electron paramagnetic resonance (EPR) experiments~\cite{spaeth03}. We
propose a new non-perturbative method that is highly superior to existing
linear response (LR) approaches~\cite{schreckenbach97,pickard02},
since it can deal with systems in which spin-orbit coupling can not be
described as a perturbation.

The total (sum of spin and orbital) magnetization can be defined from
the derivative energy $E_\mathrm{tot}$  with respect to the magnetic field $\bf B$
\beq
  \Mvec_\mathrm{} \equiv  
    \left.-\frac{\partial E_{\rm tot}}
    {\partial\Bvec}\right|_{B=0} = 
\sum_n f_n
     \left\langle\psi_n\left|-\frac{\partial\Hcal}{\partial\Bvec}\right|\psi_n
     \right\rangle_{B=0}
  \label{eq:hellmann-feynman}
\eeq
where $f_n$ is the occupation of the eigenstate $n$ and in the most
general case the expectation value is to be taken on ground state spinors
$\psi_n$. In the last equality we take advantage of the Hellmann-Feynman
theorem.  The Hamiltonian 
in atomic units is
\bea
  \Hcal_\mathrm{} &=& \frac{1}{2}\left[\pvec+\alpha\Avec(\rvec)\right]^2 +
    V(\rvec) + \nn
  &+&\frac{\alpha^2 g'}{8} \sigmavec\cdot \left[\nabla V(\rvec) \times 
    \left(\pvec+\alpha\Avec(\rvec)\right) \right],
  \label{eq:H_AE}
\eea
where we drop the trivial spin-Zeeman term,
reducing the magnetization according \eq{eq:hellmann-feynman} only
to its orbital part. We use the symmetric
gauge $\Avec(r)=\frac{1}{2}{\Bvec}\times {\rvec}$.  The last term in
\eq{eq:H_AE} is the leading spin-orbit term, describing the on-site SO
coupling (with fine structure constant $\alpha=1/c$ and the abbreviation
$g'=2(g_e-1)$~\cite{schreckenbach97,pickard02}) and $\sigmavec$ are
the Pauli matrices.  We neglect the spin other orbit (SOO) term, in
general a small contribution to the orbital magnetization and to the
$g$-tensor~\cite{patchkovskii05}.  

By inserting Eq. (\ref{eq:H_AE}) in Eq. (\ref{eq:hellmann-feynman})
we obtain:
\beq
  \Mvec =
    \frac{\alpha}{2}\sum_n f_n\Braket{\psi_n|\rvec\times\vvec|\psi_n}
    \label{eq:M_vel},
\eeq
where $\vvec = -i[\rvec,\Hcal]$, with $\Hcal$ and $\psi$ computed at
$\Bvec = 0$. This expression can be directly evaluated in a finite
system, but not in extended systems because of the unboundedness of
the position operator and of the contribution of itinerant surface
currents~\cite{orbmagn}. However, in periodic systems and in the
thermodynamic limit, Eq.~(\ref{eq:M_vel}) can rewritten as a bulk
property~\cite{orbmagn,orbmagn-others}:
\bea
  \Mvec&=& 
  -\frac{\alpha N_{\rm c}}{2N_k}\mathrm{Im}
  \sum_{n\kvec}f_{n\kvec}\times\nn 
   & &
   \Bra{ \partial_{\kvec} u_{n\kvec} }
  \times(\Hcal_{\kvec} + \epsilon_{n\kvec} - 2 \epsilon_\mathrm{F})
  \Ket{ \partial_{\kvec} u_{n\kvec} }
  \label{eq:M_periodic}
\eea
where $\Hcal_{\kvec}$ is the crystal Hamiltonian with $\Bvec=0$,
$\epsilon_{n\kvec}$ and $ u_{n\kvec}$ are its eigenvalues and
eigenvectors, $\epsilon_\mathrm{F}$ is the Fermi level, $N_{\rm c}$
is the number of cells in the  system and $N_k$ the number of $k$-points.

\eq{eq:M_vel} and \eq{eq:M_periodic} are valid at an all-electron (AE)
level.  To compute the orbital magnetization within a pseudopotential
(PS) approach, we recall that a PS Hamiltonian ($\bar{\Hcal}$) reproduces
by construction differences and derivatives of the total energy. Thus we
can still obtain $\Mvec$, from \eq{eq:hellmann-feynman},  if we replace
$\partial{\Hcal}/\partial{\Bvec}$ and $\psi_n$ by the corresponding
PS quantities  $\partial{\overline{\Hcal}}/\partial{\Bvec}$ and
$\overline{\psi}_n$.

We obtain the PS Hamiltonian in presence of spin-orbit coupling
and uniform magnetic field with the Gauge Including Projector
Augmented Waves (GIPAW) method~\cite{pickard01-03}. In particular
$\overline{\Hcal}=\mathcal{T}_B^+\Hcal\mathcal{T}_B$, where $\Hcal$ is given
by \eq{eq:H_AE} and $\mathcal{T}_B$ is the GIPAW transformation [Eq. (16)
of Ref.~\onlinecite{pickard01-03}]. If the AE and PS partial waves have
the same norm the GIPAW hamitonian 
is given by
\beq
  \overline{\Hcal}
     = \overline{\Hcal}^{(0)} + \overline{\Hcal}_\mathrm{SO}^{(0)} +
     \overline{\Hcal}^{(1)} + \overline{\Hcal}_\mathrm{SO}^{(1)}+O(B^2)
     \nonumber \label{eq:H_PS}
\eeq
where
\vspace*{-0.2cm}
\beq
  \overline{\Hcal}^{(0)} = \frac{1}{2}\pvec^2 +
    V_\mathrm{ps}(\rvec) + \VRNL  \label{eq:gipaw1}
\eeq
\beq
  \overline{\Hcal}_\mathrm{SO}^{(0)} =
    \frac{g'}{8} \alpha^2\, \left[\sigmavec\cdot 
     \left(\nabla V_\mathrm{ps}(\rvec)\times\pvec\right) 
   + \sum_{\Rvec}\FRNL \right]   \label{eq:gipaw2}
\eeq
\beq
  \overline{\Hcal}^{(1)} = \frac{\alpha}{2}\Bvec\cdot
    \left( \Lvec + \sum_{\Rvec}\Rvec\times\frac{1}{i}
    \left[\rvec,\VRNL\right]\right)  \label{eq:gipaw3}
\eeq
\bea
  \overline{\Hcal}_\mathrm{SO}^{(1)} &=& \frac{g'}{16}\alpha^3\,
    \Bvec\cdot\Bigg(
    \rvec\times(\sigmavec\times\nabla V_\mathrm{ps}) + \sum_{\Rvec}\ERNL + \nn
    &+& \left. \sum_{\Rvec}\Rvec\times\frac{1}{i}
    \left[\rvec,\FRNL\right]\right).  \label{eq:gipaw4}
\eea
Here $V_\mathrm{ps}$ and $\VRNL$ are the local part, and the non-local part
in separable form of the norm-conserving PS. $\FRNL$ and $\ERNL$ are
the separable non-local GIPAW projectors, accounting respectively for
the so-called paramagnetic and diamagnetic contributions~\cite{note2}
of the atomic site $\Rvec$.

Inserting $\overline{\Hcal}^{(1)} + \overline{\Hcal}_\mathrm{SO}^{(1)}$
in \eq{eq:hellmann-feynman} we obtain:
\bea
  \Mvec &=& \Mvec_\mathrm{bare}+
     \Delta\Mvec_\mathrm{bare} + \Delta\Mvec_\mathrm{para} +
     \Delta\Mvec_\mathrm{dia}
     \label{eq:M4} \\
\Mvec_\mathrm{bare} &=& \frac{\alpha}{2} \sum_{\Rvec} \Braket{
    \rvec\times \frac{1}{i}\left[\rvec,\bar{\Hcal}^{(0)}+\bar{\Hcal}^{(0)}_{\rm SO}\right]}
    \label{eq:M_bare}\\
  \Delta\Mvec_\mathrm{bare} &=& \frac{\alpha}{2} \sum_{\Rvec} \Braket{
    (\Rvec-\rvec)\times\frac{1}{i}\left[\rvec-\Rvec,\VRNL\right]}
    \label{eq:deltaM_bare}\\
  \Delta\Mvec_\mathrm{para} &=& \frac{g'\alpha^3}{16} \sum_{\Rvec} \Braket{
    (\Rvec-\rvec)\times\frac{1}{i}\left[\rvec-\Rvec,\FRNL\right]}
    \label{eq:deltaM_para}\\
  \Delta\Mvec_\mathrm{dia} &=& \frac{g'\alpha^3}{16} \sum_{\Rvec} \Braket{\ERNL}
    \label{eq:deltaM_dia},
\eea
where $\Braket{...}$ stands for $\sum_{n\kvec} f_{n\kvec}
\Braket{\overline{u}_{n\kvec}|...|\overline{u}_{n\kvec}}$.

In a periodic system $\Mvec_\mathrm{bare}$ can be nicely calculated by
evaluating \eq{eq:M_periodic} for the GIPAW Hamiltonian $\overline{\Hcal}$
and corresponding PS eigenvectors $\overline{u}_{n\kvec}$ and
eigenvalues $\overline{\epsilon}_{n\kvec}$.  All the reconstruction
terms, Eqs.~(\ref{eq:deltaM_bare}--\ref{eq:deltaM_dia}), can be easily
evaluated in extended systems, since the non-local operators $\VRNL,
\FRNL$ and $\ERNL$ act only inside finite spherical regions, centered
around each atom.

The approach presented so far allows the calculation of the
orbital magnetization in a general PS scheme including non-collinear
spin-polarization.  In this work for the sake of simplicity we use a
collinear implementation.  All expectation values are evaluated by
assuming decoupled spin channels along the spin direction $\evec$.
In particular all the spinors are eigenvectors of $\sigmavec\cdot
\evec$ and the local and total spin ($\Svec=S\,\evec$) are aligned along
$\evec$.  Since the choice of $\evec$ changes the spin-orbit coupling,
the orbital magnetization is a function of $\evec$.  In ferromagnets,
each spin-direction $\evec$ is characterized by a corresponding total
energy, whereby the minimum of the total energy with respect to $\evec$
defines the preferred direction of the spin-alignment, the so-called
{\em easy axis} of the ferromagnet.

We implemented our method in the {\sc Quantum}-Espresso
plane wave code~\cite{espresso}. We use standard norm-conserving
pseudopotentials~\cite{TM} with two GIPAW projectors per angular momentum
channel.  Using spin polarized LDA~\cite{LDA-PZ} and PBE~\cite{PBE}
functionals, we perform standard SCF calculations including the SO
term of \eq{eq:gipaw2} in the collinear approximation within the
Hamiltonian. Then we evaluate the orbital magnetization, according
to the Eqs.  (\ref{eq:M4},\ref{eq:deltaM_bare}--\ref{eq:deltaM_dia})
and \eq{eq:M_periodic} for $\Mvec_\mathrm{bare}$.  We neglect any
explicit dependence of the exchange-correlation functional on the
current density. In practice, spin-current density-funntional theory
(SCDFT) calculations have shown to produce negligible corrections to
the orbital magnetization~\cite{sharma07}.  We compute $\Mvec(\evec)$
with $\evec$ along easy axis and along other selected directions.
The $\kvec$-derivative of the Bloch wave functions can be accurately
evaluated by either a covariant finite difference formula~\cite{sai02}
or by the $k \cdot p$ method~\cite{kdotp}.  For insulating systems both
methods provide exactly the same results; for metallic systems the
covariant derivative is more involved and we apply just the $k \cdot
p$ method.

For the ferromagnetic Fe, Co and Ni, calculations are carried out at the
experimental lattice constants.  We consider $4s$ and $4d$ states in the
valence with non-linear core-correction. 
use a relatively low cutoff of 90 Ry.  In the case of Fe, the results
do not change by more than 1\% by including $3s$ and $3p$ in valence
and working at 120 Ry.  We use a Marzari-Vanderbilt cold smearing of
0.01 Ry. We carefully test our calculations for $k$-point convergence.
In all cases  a 28$\times$28$\times$28 mesh yields converged results
within $\pm 0.0001 \mu_B$.

\begin{table}[ttt]
\vspace*{-0.2cm}
\caption{Orbital magnetic moments $\Mvec(\evec)\cdot\evec$ in $\mu_B$ per 
atom of ferromagnetic metals parallel to the spin, for different spin 
orientations $\bm{e}$. $*$~denotes the experimental easy axis.  
The interstitial contribution is defined by the difference between
$\Mvec(\evec)$ and $\Mvec_\mathrm{orb}^{MT}
= \sum_{n\kvec}\int_{\Omega_s}u_{n\kvec}^\star(\rvec)\,
\rvec\times(-i\nabla+\kvec) u_{n\kvec}(\rvec)\, d\rvec$ where $\Omega_s$
is a atom-centered sphere of radius $R_\mathrm{MT}$=2.0 $r_{\rm bohr}$.
All theoretical values are based on the gradient corrected PBE functional.
The decomposition of $\Mvec$ according to \eq{eq:M4} is shown Tab.~II of the
auxiliary material.}
\label{tab:metals}
\vspace*{-0.3cm}
\begin{center}
\begin{tabular}{l@{\hspace{5mm}}l|c|ccc|c}
\hline\hline 
Metal & $\bm{e}$ & Expt. & \multicolumn{3}{c|}{This method} & FLAPW \\
      & & \cite{meyer61} & Total & Interstitial & MT  & \cite{wu01} \\
\hline
\emph{bcc}-Fe & $[001]^*$ & 0.081 & 0.0658 & 0.0225 & 0.0433 & 0.045  \\
\emph{bcc}-Fe & $[111]$ & $-$     & 0.0660 & 0.0216 & 0.0444 &   $-$  \\
\emph{fcc}-Co & $[111]^*$ & 0.120 & 0.0756 & 0.0122 & 0.0634 & 0.073  \\
\emph{fcc}-Co & $[001]$ & $-$     & 0.0660 & 0.0064 & 0.0596 &   $-$  \\
\emph{hcp}-Co & $[001]^*$ & 0.133 & 0.0957 & 0.0089 & 0.0868 &   $-$  \\
\emph{hcp}-Co & $[100]$ & $-$     & 0.0867 & 0.0068 & 0.0799 &   $-$  \\
\emph{fcc}-Ni & $[111]^*$ & 0.053 & 0.0519 & 0.0008 & 0.0511 & 0.050  \\
\emph{fcc}-Ni & $[001]$ & $-$     & 0.0556 & 0.0047 & 0.0509 &   $-$  \\
\hline\hline
\end{tabular}
\end{center}
\vspace*{-0.6cm}
\end{table}

Tab.~\ref{tab:metals} reports our results for the orbital magnetization
of the three metals Fe, Co and Ni, together with experimental
values and a recent calculation performed by FLAPW~\cite{wu01}. All
theoretical data was obtained using the PBE functional. LDA gives
within $\pm 0.003 \mu_B$ the same values (see Tab.~II in the
additional material~\cite{extra_table}).  In order to evaluate the
contribution of the interstitial regions neglected so far in the
literature (as in ~\cite{wu01,sharma07}), we have equally computed
$(\alpha/2)\braket{\Lvec}$ only inside atomic spheres.  Except for
\emph{fcc}-Co our results agree very well with FLAPW calculations.
For Ni the influence of  contributions is indeed negligible, explaining
the agreement of early DFT calculations in this case.  For the other
ferromagnets however it becomes evident from Tab.~\ref{tab:metals}
that these contributions can by no means be neglected. For Fe e.g. the
interstitial contribution is about 50\% of that inside a MT sphere and
thus leads to considerably improved {\em ab-initio} values.
This result indicates the importance of the contributions from the
interstitial regions when benchmarking and/or developing improved DFT
functionals for orbital magnetism.

In the following we will show that the anisotropies in the orbital
magnetizations are well described to allow us to calculate the electronic
$g$-tensor of paramagnetic systems, in order to understand the microscopic
structure of radicals or paramagnetic defects in solids.  From the
orbital magnetization we can obtain the deviation of the $g$-tensor,
$\Delta g_{\mu\nu}$ from the free electron value $g_e$=2.002319 by the
variation of $\Mvec$ with a spin flip:
\beq
   \Delta g_{\mu\nu} 
     = -\frac{2}{\alpha}\evec_\mu\cdot
     \frac{\Mvec(\evec_\nu)-\Mvec(-\evec_\nu)}{ S - (-S) } 
   = -\frac{2}{\alpha S}\evec_\mu\cdot\Mvec(\evec_\nu)
 \label{eq:g-tensor}
\eeq
where $\nu$, $\mu$ are Cartesian directions of the magnetic field and the
total spin $S$, respectively.  To get the full tensor $\Delta g_{\mu\nu}$,
for every paramagnetic systems we carry out three calculations by aligning
the spin quantization axis along the three Cartesian directions.

\begin{table}[ttt]
\vspace*{-0.2cm}
\caption{Principal values $\Delta g$ in ppm for the diatomic molecules of the
RnF-family calculated by linear response (LR)~\cite{pickard02} and
with the current method. $\parallel$ is symmetry axis of the dimer.
$\Delta g (\Delta M)$ gives the contributions
of $\Delta\Mvec_\mathrm{bare}$, $\Delta\Mvec_\mathrm{para}$,
$\Delta\Mvec_\mathrm{dia}$ to the $g$-tensor. A (small) relativistic mass
correction term $\Delta g_\mathrm{RMC}$~\cite{pickard02} is included in
both sets of data.}
\label{tab:RnF-family}
\vspace*{-0.3cm}
\begin{center}
\begin{tabular}{llrrrr}
\hline\hline
  & & Linear response & This method & \hspace{4mm}$\Delta g (\Delta M)$  \\
\hline
NeF         & $\Delta g_\parallel$ &     $-$336 &    $-$328 &   $-$414 \\
            & $\Delta g_\perp$     &      52633 &     52778 &     2935 \\
ArF         & $\Delta g_\parallel$ &     $-$349 &    $-$343 &  $-$4450 \\
            & $\Delta g_\perp$     &      42439 &     42519 &     2914 \\
KrF         & $\Delta g_\parallel$ &     $-$360 &    $-$353 &   $-$968 \\
            & $\Delta g_\perp$     &      59920 &     59674 &  $-$1918 \\
XeF         & $\Delta g_\parallel$ &     $-$358 &    $-$354 &  $-$3733 \\
            & $\Delta g_\perp$     &     163369 &    158190 & $-$55099 \\
RnF         & $\Delta g_\parallel$ &     $-$356 &    $-$299 & $-$13670 \\
            & $\Delta g_\perp$     &     603082 &    488594 &$-$255079 \\
\hline\hline
\end{tabular}\end{center}
\vspace*{-0.5cm}
\end{table}

\begin{table}[bbb]
\vspace*{-0.5cm}
\caption{Calculated principal values $\Delta g$ in ppm for the diatomic molecules
of the PbF-family. See Tab.~\ref{tab:RnF-family} for details.}
\label{tab:PbF-family}
\vspace*{-0.3cm}
\begin{center}
\begin{tabular}{llrrrr}
\hline\hline
   & & Linear Response & This method & \hspace{4mm}$\Delta g (\Delta M)$  \\
\hline
CF   & $\Delta g_{\parallel}$ &   $-\infty$ & $-$1999719 & $-$119746 \\ 
     & $\Delta g_{\perp}$     &        1920 &     $-$553 &    $-$240 \\
SiF  & $\Delta g_{\parallel}$ &   $-\infty$ & $-$1995202 & $-$100021 \\ 
     & $\Delta g_{\perp}$     &      $-$480 &    $-$2470 &    $-$535 \\
GeF  & $\Delta g_{\parallel}$ &   $-\infty$ & $-$1998078 &  $-$40609 \\ 
     & $\Delta g_{\perp}$     &    $-$15505 &   $-$39101 &    $-$388 \\
SnF  & $\Delta g_{\parallel}$ &   $-\infty$ & $-$1996561 &  $-$72464 \\ 
     & $\Delta g_{\perp}$     &    $-$64997 &  $-$142687 &   $-$5339 \\
PbF  & $\Delta g_{\parallel}$ &   $-\infty$ & $-$1999244 &  $-$90214 \\ 
     & $\Delta g_{\perp}$     &   $-$288383 &  $-$556326 &  $-$22476 \\
\hline\hline
\end{tabular}\end{center}
\vspace*{-0.4cm}
\end{table}

To evaluate the approach, we compute the $g$-tensors of selected
diatomic radicals. An energy cutoff of 100~Ry is used in all
molecular calculations. They are performed in a cubic repeated
cell with a large volume of 8000~\AA$^3$ and the Brillouin zone is
sampled only at the $\Gamma$ point.  For comparison, we also compute
the $g$-tensor via the linear response method (LR)~\cite{pickard02},
which we recently implemented in the {\sc Quantum}-Espresso package.
For a wide range of molecular radicals including almost all of
the examples discussed in Ref.~\cite{pickard02} the new approach
reproduces the values obtained via LR within a few ppm (see also
auxiliary Tab.~I in~\cite{extra_table}). In Tab.~\ref{tab:RnF-family}
and~\ref{tab:PbF-family} we report the calculated  principal components
of the computed $g$-tensors for the RnF and PbF families.  For the
members of the RnF family qualitative deviations are only observed
if heavy elements like Xe and Rn are involved, showing in LR large
deviations $\Delta g_{\perp}$ of up to $10^5$~ppm from $g_e$ for the
corresponding fluorides. The treatment of SO-coupling beyond LR leads
to considerably smaller values of $\Delta g_{\perp}$, reduced  by 3\%
(XeF) and 19\% (RnF), respectively.  Note that the reconstruction terms,
Eqs.~(\ref{eq:deltaM_bare}--\ref{eq:deltaM_dia}), significantly contribute
to the $g$-tensor. For the RnF family (see Tab.~\ref{tab:RnF-family})
this is essential to obtain a value of $\Delta g_\parallel \approx
0$~\cite{dimitriev92} as also expected analytically~\cite{note4}.

In contrast to the RnF family (5 electrons in the $p$-shell, $e^4 a_1^1$
electronic configuration), the PbF family has only one electron within
the $p$-shell. Without SO-coupling the unpaired electron occupies a
degenerate $e$-level. Consequently, without SO, the HOMO-LUMO gap between
the unpaired electron and the empty levels is zero, leading within LR
to diverging values $g_{\parallel}$.  This failure of LR is observed for
all members of the PbF family, already for CF containing light elements
exclusively.  In contrast, our new method circumvents perturbation theory,
and predicts a nearly vanishing $g$-value $g_{\parallel}=g_e+\Delta
g_{\parallel}\approx 0$ along the bond direction of the diatomic molecules
as expected analytically~\cite{note4}.

In conclusion, we have shown how a recently developed formula
for the orbital magnetization can be applied in an {\em ab-initio}
pseudopotential scheme whereby the spin-orbit coupling enters explicitly
the self-consistent cycle.  In comparison with linear response methods,
our approach allows an improved calculation of the electronic $g$-tensor
of paramagnetic systems containing heavy elements or with large
deviations of the $g$-tensor from the free electron value. The latter
situation is encountered in many paramagnetic centers in solids, such
as those exhibiting a Jahn-Teller distortion~\footnote{ The new method
is e.g. applicable in case of some intrinsic defects (silicon anti-sites)
in the compound semiconductor SiC where LR fails to achieve convergence
with respect to $k$-points; U. Gerstmann {\em et al.}, unpublished.}
and/or containing transition metal impurities.  In addition, our method
provides improved orbital magnetizations with respect to the preexisting
approaches that neglect the contributions of the interstitial regions.
This has been shown for the highly ordered ferromagnets where the orbital
contribution is partially quenched by the crystal field.  The presented
approach is perfectly suited to describe also the ferromagnetism of
nanostructures where the orbital quench is weaker and the orbital part
of the magnetic moments becomes more dominant.

U.~G. acknowledges financial support by the DFG (Grant No. GE 1260/3-1)
and by the CNRS. D.~C. acknowledges partial support from ENI.
Calculations were performed at the IDRIS, Paris (Grant No. 061202)
and at CINECA, Bologna (Grant Supercalcolo 589046187069).


\newpage

\begin{widetext}
In this auxiliary material we report (1) the calculated EPR $g$-tensor
for a set of molecular radicals including almost all of the examples
discussed in Ref.~\cite{pickard02}; (2) the decomposition of the orbital
magnetization of Fe, Co and Ni according to Eq.~(9); (3) the orbital
magnetization of Fe, Co and Ni, integrated within atomic spheres.
\vspace{1cm}

\begin{center}
\begin{tabular}{ll cr rrr}
\hline\hline
radical & & LR & This method & $g(M')$ & $g(\Delta M$)& $\Delta g_{\rm RMC}$\\
\hline
H$_2^+$     & $\Delta g_\parallel$ &    $-$39.3 &   $-$39.3 &      24.7 &      0 &     -64\\
            & $\Delta g_\perp$     &    $-$41.7 &   $-$41.7 &      22.3 &      0 &     -64\\
CN          & $\Delta g_\parallel$ &     $-$141 &    $-$139 &        32 &      9 &    -180\\
            & $\Delta g_\perp$     &    $-$2600 &   $-$2603 &   $-$2192 & $-$231 &    -180\\
CO$^+$      & $\Delta g_\parallel$ &     $-$136 &    $-$134 &        12 &     33 &    -179\\
            & $\Delta g_\perp$     &    $-$3229 &   $-$3231 &   $-$3052 & $-$260 &    -179\\
BO          & $\Delta g_\parallel$ &      $-$70 &     $-$75 &      $-$5 &     22 &     -92\\
            & $\Delta g_\perp$     &    $-$2384 &   $-$2384 &   $-$2163 & $-$129 &     -92\\
BS          & $\Delta g_\parallel$ &      $-$81 &     $-$82 &    $-$154 &    177 &    -105\\
            & $\Delta g_\perp$     &    $-$9990 &  $-$10001 &   $-$9513 & $-$382 &    -105\\
AlO         & $\Delta g_\parallel$ &     $-$149 &    $-$149 &       339 & $-$294 &    -192\\
            & $\Delta g_\perp$     &    $-$1834 &   $-$1842 &   $-$1316 & $-$334 &    -192\\
NeF         & $\Delta g_\parallel$ &     $-$336 &    $-$328 &        86 &      6 &    -420\\
            & $\Delta g_\perp$     &      52633 &     52778 &     49843 &   3355 &    -420\\
MgF         & $\Delta g_\parallel$ &      $-$59 &     $-$68 &        57 &  $-$37 &     -88\\
            & $\Delta g_\perp$     &    $-$2283 &   $-$2316 &   $-$2227 &   $-$1 &     -88\\
ArF         & $\Delta g_\parallel$ &     $-$349 &    $-$343 &       102 &  $-$10 &    -435\\
            & $\Delta g_\perp$     &      42439 &     42519 &     39605 &   3349 &    -435\\
KrF         & $\Delta g_\parallel$ &     $-$360 &    $-$353 &       615 &  $-$520 &   -448\\
            & $\Delta g_\perp$     &      59920 &     59674 &     61593 & $-$1470 &   -448\\
XeF         & $\Delta g_\parallel$ &     $-$358 &    $-$354 &      3380 & $-$3283 &   -450\\
            & $\Delta g_\perp$     &     163369 &    158190 &    213285 & $-$54649 &  -450\\
HgF         & $\Delta g_\parallel$ &     $-$288 &    $-$263 &     54601 & $-$54490 &  -374\\
            & $\Delta g_\perp$     &   $-$34268 &  $-$33355 &     52161 & $-$85115 &  -374\\
RnF         & $\Delta g_\parallel$ &     $-$356 &    $-$299 &     13371 & $-$13196 &  -474\\
            & $\Delta g_\perp$     &     603082 &    488594 &    743638 & $-$254605 & -474\\
\hline\hline
\end{tabular}
\end{center}
TABLE I: calculated $\Delta\tensor{g}$ in ppm for diatomic molecules, by
linear response (LR)~\cite{pickard02} and with the current method. For
sake of comparison, the SOO contribution is omitted from the GIPAW
results.  The ``$\Delta$ contrib.'' column contains the sum of the
contributions of $\Delta\Mvec_\mathrm{bare}$, $\Delta\Mvec_\mathrm{para}$
and $\Delta\Mvec_\mathrm{dia}$ to the $g$-tensor. The relativistic mass
correction term $\Delta g_{RMC}$ included in both sets of data is given
explicitly.
\vspace{2cm}

\begin{center}
\begin{tabular}{lr @{\hspace{1.0cm}} r @{\hspace{0.5cm}} r 
                   @{\hspace{0.5cm}} r @{\hspace{0.5cm}} r}
\hline\hline
Metal &  & $M_\mathrm{bare}$ & $\Delta M_\mathrm{bare}$
         & $\Delta M_\mathrm{para}$ & $\Delta M_\mathrm{dia}$\\
\hline
\emph{bcc}-Fe & LDA & 0.0616 & 0.0005 & 0.0016 & 0.0003 \\
              & PBE & 0.0639 & 0.0000 & 0.0016 & 0.0003 \\
\emph{fcc}-Co & LDA & 0.0706 & 0.0019 & 0.0014 & 0.0002 \\
              & PBE & 0.0722 & 0.0018 & 0.0014 & 0.0002 \\
\emph{hcp}-Co & LDA & 0.0875 & 0.0032 & 0.0014 & 0.0003 \\
              & PBE & 0.0908 & 0.0032 & 0.0014 & 0.0003 \\
\emph{fcc}-Ni & LDA & 0.0519 & 0.0019 & 0.0007 & 0.0000 \\
              & PBE & 0.0494 & 0.0017 & 0.0007 & 0.0001 \\
\hline\hline
\end{tabular}
\end{center}
TABLE II: Contributions to the orbital magnetization along the easy axis,
in $\mu_\mathrm{B}$ per atom. See eq.~(9) in the text. As in the case of
molecules, the ``$\Delta$ contrib.'' is not negligible and it is comparable
to the difference between the full orbital magnetization and the orbital
magnetization calculated inside atomic spheres (see Tab.~III in this
auxiliary material.
\vspace{2cm}

\begin{center}
\begin{tabular}{l @{\hspace{1.0cm}} rr @{\hspace{1cm}} rr }
\hline\hline
Metal & FLAPW LDA~\cite{wu01} & FLAPW PBE~\cite{wu01} 
      & This work LDA & This work PBE\\
\hline
\emph{bcc}-Fe & 0.048 & 0.045  & 0.0452 & 0.0433 \\
\emph{fcc}-Co & 0.076 & 0.073  & 0.0641 & 0.0634 \\
\emph{hcp}-Co & $-$   & $-$    & 0.0835 & 0.0868 \\
\emph{fcc}-Ni & 0.049 & 0.050  & 0.0499 & 0.0511 \\
\hline\hline
\end{tabular}
\end{center}
TABLE III: Orbital magnetization contribution inside
atomic spheres, in $\mu_\mathrm{B}$ per atom, along the
easy axis.  This is defined as $\Mvec_\mathrm{orb}^s
= \sum_{n\kvec}\int_{\Omega_s}u_{n\kvec}^\star(\rvec)\,
\rvec\times(-i\nabla+\kvec) u_{n\kvec}(\rvec)\, d\rvec$ where $\Omega_s$
is a sphere centered on one atom, of radius $R_\mathrm{MT}$.
$R_\mathrm{MT}$ is given in units of the Bohr radius ($a_0$).
$R_\mathrm{MT} = 2.0\,a_0$ is a typical muffin-tin radius used in FLAPW
calculations for Fe, Co and Ni. Our results agree very well with FLAPW
results. By comparing to the orbital magnetization calculated according
to the periodic formula (Tab.~III of the paper), which takes into
account not only the atomic spheres but also the interstitial region,
it is evident that contribution from the interstitial is not negligible.
\end{widetext}


\vspace*{-0.3cm}
\begin{thebibliography}{99}

\vspace*{-0.4cm}

\bibitem{MTP}
  R.D. King-Smith and D. Vanderbilt,
  Phys. Rev. B \textbf{47}, 1651 (1993); 
  D. Vanderbilt and R.D. King-Smith,
  Phys. Rev. B \textbf{48}, 4442 (1993).

\bibitem{resta-RMP}
  R. Resta, Rev. Mod. Phys. \textbf{66}, 899 (1994).

\bibitem{orbmagn}
  T. Thonhauser, D. Ceresoli, D. Vanderbilt, and R. Resta,
  Phys. Rev. Lett. \textbf{95}, 137205 (2005);
  D. Ceresoli, T. Thonhauser, D. Vanderbilt, and R. Resta,
  Phys. Rev. B \textbf{74}, 024408 (2006).

\bibitem{orbmagn-others}
  D. Xiao, J. Shi, and Q. Niu,
  Phys. Rev. Lett. \textbf{95}, 137204 (2005);
  J. Shi, G. Vignale, D. Xiao, and Q. Niu,
  Phys. Rev. Lett. \textbf{99}, 197202 (2007).

\bibitem{wu01}
  R.~Wu, \emph{First Principles Determination of Magnetic Aniso\-tropy and
  Magnetostriction in Transition Metal Alloys}, Lecture Notes in Physics {\bf 580}, 
  Springer, Berlin (2001).

\bibitem{sharma07}
  S. Sharma \emph{et al.}, Phys. Rev. B \textbf{76}, 100401 (2007).

\bibitem{trygg95}
  O. Eriksson, M. S. S. Brooks, and B. Johansson,
  Phys. Rev. B \textbf{41}, 7311 (1990);
  J. Trygg, B. Johansson, O. Eriksson, and J.M. Wills,
  Phys. Rev. Lett. \textbf{75}, 2871 (1995).

\bibitem{spaeth03} 
  J.-M. Spaeth and H. Overhof,
  {\em Point defects in semiconductors and insulators}, 
  Springer Berlin (2003).

\bibitem{schreckenbach97}
  G. Schreckenbach and T. Ziegler,
  J. Phys. Chem. A \textbf{101}, 3388 (1997).

\bibitem{pickard02}
  C. J. Pickard, F. Mauri,
  Phys. Rev. Lett. \textbf{88}, 086403 (2002).

\bibitem{patchkovskii05}
  S. Patchkovskii, R.T. Strong, C.J. Pickard and S. Un,
  J. Chem. Phys. \textbf{122}, 214101 (2005).

\bibitem{pickard01-03}
  C. J. Pickard and F. Mauri, 
  Phys. Rev. B \textbf{63}, 245101 (2001);
  C. J. Pickard and F. Mauri,
  Phys. Rev. Lett. \textbf{91}, 196401 (2003).

\bibitem{note2}
Given the set of GIPAW projectors $\ket{\ptilde_{\Rvec,n}}$, the
diamagnetic and the paramagnetic term are $\ERNL=\sum_{\Rvec,nm}
\ket{\ptilde_{\Rvec,n}}{\bf e}_{\Rvec,nm}\bra{\ptilde_{\Rvec,m}}$
and $\FRNL=\sum_{\Rvec,nm}\ket{\ptilde_{\Rvec,n}}\sigmavec\cdot
\bm{f}_{\Rvec,nm} \bra{\ptilde_{\Rvec,m}}$. The expression of
${\bf e}_{\Rvec,nm}$ and $\bm{f}_{\Rvec,nm}$, respectively, is given by
Eq.~(11) and Eq.~(10) of Ref.~\cite{pickard02}.

\bibitem{espresso}
  P. Giannozzi \emph{et al.},
    J. Phys.: Condens. Matter {\bf 21}, 395502 (2009);
    http://www.quantum-espresso.org

\bibitem{TM} N. Troullier, J. L. Martins, Phys. Rev. B {\bf 43},
   1993 (1991).

\bibitem{LDA-PZ}
   J.P. Perdew, A. Zunger, Phys. Rev. B {\bf 23}, 5048 (1981).

\bibitem{PBE}
  J.P. Perdew, K. Burke, and M. Ernzerhof,
  Phys. Rev. Lett. \textbf{78}, 1396 (1997).

\bibitem{sai02}
  N. Sai, K.M. Rabe, and D. Vanderbilt,
  Phys. Rev. B \textbf{66}, 104108 (2002).

\bibitem{kdotp}
  C.J . Pickard and M. C. Payne, Phys. Rev. B \textbf{62}, 4383 (2000);
  M. Iannuzzi and M. Parrinello, Phys. Rev. B \textbf{64}, 233104 (2001).

\bibitem{meyer61}
  A.J.P. Meyer and G. Asch,
  J. Appl. Phys \textbf{32}, S330 (1961).

\bibitem{extra_table}
  See EPAPS Document No. XXXX. For more information, see
  \url{http://www.aip.org/pubservs/epaps.html}.

\bibitem{dimitriev92}
 Y. Y. Dmitriev \emph{et al.}, Phys. Lett. A \textbf{167}, 280 (1992).

\bibitem{note4}
Along the bond direction (in absence of SO) the angular momentum is a
good quantum number leading to $g_{\parallel}\approx g_e + 2 m_l$: for
the PbF family $g_{\parallel}\approx 0$ (the unpaired electron occupies
the $m_l=-1$ $p$-like orbital); for the RnF family $m_l$=$0$ and, thus,
$g_{\parallel}\approx g_e$.

\end{thebibliography}
\end{document}